\documentclass[useAMS,usenatbib,usegraphicx]{mn2e}

\title[]{
Contrasting the chemical evolution of the Milky Way and Andromeda galaxies
}
\author[Agostino Renda, Daisuke Kawata, Yeshe Fenner, Brad K. Gibson]{Agostino Renda$^{1}$\thanks{E-mail:
arenda,dkawata,yfenner,bgibson@astro.swin.edu.au
}, Daisuke Kawata$^{1}$\footnotemark[1], Yeshe Fenner$^{1}$\footnotemark[1],
Brad K. Gibson$^{1}$\footnotemark[1]\\ $^{1}$Centre for Astrophysics \& Supercomputing, Swinburne University, Hawthorn, Victoria, 3122, Australia\\
}
\begin{document}

\date{Accepted. Received; in original form}

\pagerange{\pageref{firstpage}--\pageref{lastpage}} \pubyear{2004}

\maketitle

\label{firstpage}

\begin{abstract}
The chemical evolution history of a galaxy hides clues about how it
formed and has been changing through time. We have studied the
chemical evolution history of the Milky Way (MW) and Andromeda (M31)
to find which are common features in the chemical evolution of disc
galaxies as well as which are galaxy-dependent. We use a semi-analytic
multi-zone chemical evolution model. Such models have succeeded in
explaining the mean trends of the observed chemical properties in 
these two Local Group spiral galaxies with similar mass and
morphology. 
Our results suggest that while the evolution of the MW and M31 shares
general similarities, differences in the formation history are
required to explain the observations in detail.
In particular, we found that the observed higher metallicity in the M31 halo
can be explained by either a) a higher halo star formation efficiency 
or b) a larger reservoir of infalling halo gas 
with a longer halo formation phase.
These two different pictures
would lead to a) a higher $[$O/Fe$]$ at low metallicities or 
b) younger stellar populations in the M31 halo, respectively. 
Both pictures result in a more massive stellar halo in M31, which
suggests a possible correlation between the halo metallicity and 
its stellar mass.
\end{abstract}

\section{Introduction} 
The chemical properties of galaxies hide
clues about their formation and evolution. 
Semi-analytic chemical evolution models
(\citealt{TA}; \citealt{Tinsley}) have succeeded in explaining the
mean trends of galactic systems by numerically solving a set of
equations governing the simplified evolution of the chemical elements
as they cycle through gas and stars.  One strength of these models is
that they typically have the fewest number of free parameters, making
convergence to a smaller set of solutions more likely.

Strong constraints can be placed on chemical evolution models only by
contrasting them against a comprehensive set of observed properties.
Since the most detailed observational data are generally 
available for the Milky Way (e.g. \citealt{FBH}), successful
agreement between model predictions and these observed properties has
been obtained by several studies in the past, which help us to
understand the formation history of the Milky Way (MW). Some models
have focused on the evolution of the chemical abundances both in the
solar neighbourhood and in the whole disc, adopting a framework in which 
the MW has been built-up inside-out by means of a single accretion event (e.g. \citealt{MF};
\citealt{PT}). Others have used an early infall of gas to explain the
thick disc formation, followed by a slower infall to form the thin
disc (e.g. Chiappini, Matteucci \& Gratton 1997). Several studies have
paid particular attention to the chemical evolution of a larger range
of elements (e.g. Timmes, Woosley \& Weaver 1995; \citealt{GP};
Alib\'es, Labay \& Canal 2001) or focused on the features of the
Metallicity Distribution Function (MDF) in the solar neighbourhood
(e.g. \citealt{FG}). However, an implicit assumption remains unanswered: 
\textit{is the MW a typical spiral?}

Andromeda (M31) is the closest spiral to our MW (e.g. \citealt{vanDenBergh2003}). 
Previous theoretical studies of the chemical
evolution of M31 have been done by \cite{DT} and Moll\'a, Ferrini \&
Diaz (1996), both emphasising the evolution of the M31 disc. They
concluded that M31 has a formation history and chemical evolution
similar to that of the MW. More recently, however, there has been
considerable observational progress in the study of both galaxies. 
A striking difference between M31 and the MW is that the
metallicity of the M31 halo ($\langle$[Fe/H]$\rangle\approx-0.5$)
\footnote{Hereafter [X/Fe]$=\log_{10}(X/Fe)-\log_{10}(X/Fe)_{\odot}$.} 
is significantly higher than
its MW analogue ($\langle$[Fe/H]$\rangle\approx -1.8$,
e.g. \citealt{RN}), as revealed by many recent studies
(e.g. Holland, Fahlman \& Richer 1996; Durrell, Harris, 
\& Pritchet 2001; \citealt{SvD}; \citealt{FJ}; \citealt{RG}; \citealt{FergusonETal};
\citealt{WE}; \citealt{BellazziniETal}; \citealt{BrownETal}; \citealt{ReitzelETal}; \citealt{RichETal}). 
 
It is therefore timely to attempt the construction
of a chemical evolution model for both the MW and M31, 
using the same framework. 
Such an attempt may be helpful in highlighting the 
common features in the chemical evolution of spiral galaxies 
(at least in these two spirals) 
and those which remain galaxy-dependent.

In Section~2 we describe our multi-zone
chemical evolution model, and in Section~3 we
present the results of our MW model. In Section~4 we show the results
for M31. Finally, our results are discussed in Section~5.

\section{The model}

In this study we use a semi-analytic multi-zone chemical evolution
model for a spiral galaxy. This model is based on \texttt{GEtool}
(\citealt{FG}; \citealt{GibsonETal}). We follow the chemical evolution of the halo and
disc of M31 and the MW under the assumption that both
galaxies are formed from two phases of gas infall. The first infall
episode corresponds to the halo build-up, and the second to the
inside-out formation of the disc. Similar formalisms have been
successful in modeling the chemical evolution of the solar
neighbourhood (e.g.  Chiappini, Matteucci \& Gratton 1997; \citealt{ChangETal}; Alib{\' e}s, Labay \& Canal 2001). We assume that halo
stars were born in a burst induced by the collapse of a single
proto-galactic cloud
(Eggen, Lynden-Bell \& Sandage 1962) or by the multiple merger of building-blocks
(\citealt{SZ}; \citealt{BC};
\citealt{BrookETal}). 
The disc is assumed to be built-up
inside-out by the smooth accretion of gas on a longer timescale.
Observations of HI High Velocity Clouds (HVCs) that appear to be
currently falling onto the MW (e.g. \citealt{PutmanETal} and
references therein) may provide evidence for such gas infall.  Recently,
HVCs have also been detected in the M31 neighbourhood, though their
interpretation as infalling clouds is debated \citep{ThilkerETal}.

We monitor the face-on projected properties of the halo and disc
components. While this geometrical simplification is suitable for
approximating the flat disc, it is less appropriate for the halo,
whose shape is roughly spherical rather than disc-like.  
However, we consider this choice acceptable in a simplified model of the 
chemical evolution of a spiral galaxy.
We follow the chemical evolution of several independent rings, $2$~kpc
wide, out to a galactocentric radius $R = 10 R_{d}$,
where $R_{d}$ is the disc scale-length. We also
ignore the bulge component, because we are interested in the
relatively outer region ($R>4$~kpc). 
Each ring is a single zone onto which gas falls, 
without exchange of matter between the rings. 
We trace the chemical evolution of each zone individually. 
In our model, we assume that the age of the galaxy
is $t_{now} = 13~Gyr$. The basic equations (e.g. \citealt{Tinsley}), 
in a zone at a radius $R$,
for the evolution of the gas surface density $\Sigma_{g,i}(R,t)$ of an
element $i$ are written as follows:
\begin{eqnarray}
&&\dot\Sigma_{g,i}(R,t)=\nonumber\\
&-&\psi(R,t)X_{i}(R,t) + \int_{M_{min}}^{M_{Bmin}}\psi(R,t-\tau_{m})\nonumber\\
&\times&Y_{i}(m,Z_{t-\tau_{m}})\frac{\varphi(m)}{m}dm + k\int_{M_{B_{min}}}^{M_{B_{max}}}\frac{\varphi(M_{B})}{M_{B}}\nonumber\\
&\times&\int_{\mu_{min}}^{0.5}f(\mu)\psi(R,t-\tau_{m_{2}})Y_{i}(M_{B},Z_{t-\tau_{m_{2}}})d\mu~dM_{B}\nonumber\\
&+& (1-k)\int_{M_{Bmin}}^{M_{Bmax}}\psi(R,t-\tau_{m})Y_{i}(m,Z_{t-\tau_{m}})\nonumber\\
&\times&\frac{\varphi(m)}{m}dm + \int_{M_{Bmax}}^{M_{max}}\psi(R,t-\tau_{m})Y_{i}(m,Z_{t-\tau_{m}})\nonumber\\
&\times&\frac{\varphi(m)}{m}dm + X_{i_{halo}}(t)h(R)e^{\frac{-t}{\tau_{h}}}\nonumber\\
&+&X_{i_{disc}}(t)d(R)e^{\frac{-(t-t_{delay})}{\tau_{d}(R)}}.\end{eqnarray}
Here $X_{i}(R,t) = \frac{\Sigma_{g,i}(R,t)}{\Sigma_{g}(R,t)}$ is the
mass fraction for the element $i$; $\psi(R,t)$ is the star formation
rate (SFR);
$\varphi(m)$ is the Initial Mass Function (IMF) with the mass range
$M_{min}$ - $M_{max}$; $\tau_{m}$ is the lifetime of a star with mass
$m$; $Y_{i}(m,Z_{t-\tau_{m}})$ is the stellar yield of the element
from a star of mass $m$, age $\tau_{m}$ and metallicity
$Z_{t-\tau_m}$. The first term describes the depletion of the element $i$ which is
locked-up in newly formed stars. The second and the fourth terms show
the contribution of mass loss from low and intermediate mass
stars. The third term describes the contribution from Type Ia SNe
(SNeIa). The contribution from SNeIa is calculated as suggested in
\cite{GR}, where $k$, $M_{B_{min}}$, $M_{B_{max}}$, $\mu_{min}$, $\mu$, $f(\mu)$, $\tau_{m_{2}}$ are defined. 
The fifth term shows the contribution from
Type II SNe (SNeII). The sixth and the seventh terms represent the infalling halo and disc gas, respectively.

The Kroupa, Tout \& Gilmore (1993) IMF is used here. We have chosen a
lower mass limit of $M_{min}=0.08$~M$_{\odot}$ and imposed an upper mass
limit of $M_{max}=60$~M$_{\odot}$ in order to avoid the overproduction of oxygen and recover the
observed trend of $[$O/Fe$]$ at low metallicity at the solar neighbourhood in the MW. 
Such IMF upper mass limit is currently 
loosely constrained by stellar formation and evolution models.
Yet, these stellar models, and the yields they provide, are one of the most
important features in galactic chemical evolution, although
questions remain concerning the precise composition of stellar ejecta,
due to the uncertain role played by processes including mass loss,
rotation, fall-back, and the location of the mass cut, which separates
the remnant from the ejected material in SNe. 
The SNeII
yields are from \cite{WW}. We have halved the
iron yields shown in \cite{WW}, as suggested by
\cite{TimmesETal}.  The SNeIa yields are from \cite{IwamotoETal}.
 The metallicity-dependent yields of \cite{RV} have
been used for stars in the range 1 - 8~M$_{\odot}$.
The lifetimes of stars as a function of mass and metallicity have been
taken from \cite{SchallerETal}.

\begin{table*}
\caption{The parameters for the MW and M31 models.}
\centering
\begin{minipage}{140mm}
\begin{tabular}{@{}lllllllllll@{}}
\hline
     &$\alpha$&$\Sigma_{t,h}(R_{\odot},t_{now})$&$R_{d}$&$\Sigma_{t,d}(R_{\odot},t_{now})$&$\tau_{h}$&$t_{delay}$&$a_{d}$&$b_{d}$&$\nu_{h}$&$\nu_{d}$\\
  & & (M$_{\rm \sun}$ pc$^{-2}$) & (kpc) & (M$_{\rm \sun}$ pc$^{-2}$) &
 (Gyr) & (Gyr) & (Gyr) & & & \\ \hline
MW&$2$&$~6$&$3.0$&$48$&$0.1$&$1.0$&$2.0$&$1.25$&$~0.125$&$0.03$\\
M31a&$2$&$~6$&$5.5$&$46$&$0.1$&$1.0$&$2.0$&$1.25$&$~0.125$&$0.03$\\
M31b&$2$&$~6$&$5.5$&$46$&$0.1$&$1.0$&$0.7$&$0.50$&$12.5~~$&$0.08$\\
M31c&$2$&$57$&$5.5$&$46$&$0.5$&$6.0$&$0.1$&$0.10$&$~0.125$&$0.08$\\
\hline
\end{tabular}
\end{minipage}
\end{table*}

\subsection{Infall}
The infall rate during the halo and disc phases is simply assumed
to decline exponentially,
as seen by the adopted sixth and seventh terms in equation (1). The evolution
of the total\footnote{Hereafter, by ``total'' density we mean the sum of the stellar and gas densities.} surface mass density,
$\Sigma_{t}(R,t)$, is described by:
\begin{eqnarray}
\frac{d\Sigma_{t}(R,t)}{dt}=h(R)e^{\frac{-t}{\tau_{h}}} + d(R)e^{\frac{-(t-t_{delay})}{\tau_{d}(R)}}.
\end{eqnarray}
Here, the first term describes the infall rate in the halo phase. 
The infall time-scale in the halo phase, $\tau_{h}$, is assumed to be 
independent of radius, for simplicity. The infall of disc gas starts 
with a delay of $t_{delay}$, as seen in the second term. 
The time-scale of the disc infall depends on radius as follows:
\begin{eqnarray}
\tau_{d}(R) = a_{d} + b_{d} \frac{R}{kpc} Gyr.
\end{eqnarray}
The values for the constants $a_{d}$ and $b_{d}$ are free parameters.
The infall coefficients $h(R)$ and $d(R)$ are chosen in order to
reproduce the present-day total surface
density in the disc, $\Sigma_{t,d}(R,t_{now})$, and halo,
$\Sigma_{t,h}(R,t_{now})$, respectively, as follows
(e.g. \citealt{TimmesETal}):
\begin{eqnarray}
h(R)&=&\Sigma_{t,h}(R,t_{now})\nonumber\\
&&\times\Bigg\{\tau_{h}\Big[1-exp\Big(\frac{-t_{now}}{\tau_{h}}\Big)\Big]\Bigg\}^{-1};\\
d(R)&=&\Sigma_{t,d}(R,t_{now})\nonumber\\
&&\times\Bigg\{\tau_{d}(R)\Big[1-exp\Big(-\frac{t_{now}-t_{delay}}{\tau_{d}(R)}\Big)\Big]\Bigg\}^{-1}.
\end{eqnarray}
The infalling halo gas has been assumed to be of primordial composition.
On the other hand, it is unlikely that the accreting gas has primordial 
abundance at a later epoch, since even low density inter-galactic medium,
such as the Lyman ${\alpha}$ forest, has a significant amount of metals
(e.g. \citealt{CS}), and it is known that the HVCs in the MW,
which may be infalling gas clouds, 
have metallicities between $0.1$ and $0.3~Z_{\odot}$
\citep{SembachETal}. Therefore, we assume that the gas accreting onto the disc is 
pre-enriched. The level of pre-enrichment can be loosely constrained
from the observed metallicity of Galactic HVCs.
We simply assume that the metallicity of the infalling disc 
material is  $Z_{infall}(R,t) = Z(R,t)$ if
$Z(R,t) < Z_{infall,max} = 0.3~Z_{\odot}$, where $Z(R,t)$ is the
metallicity of the gas at the radius $R$ and the time $t$,
otherwise $Z_{infall}(R,t) = Z_{infall,max}=0.3~Z_{\odot}$.
The abundance pattern of the infalling disc gas 
is further unknown parameter. Following the above simple assumption,
we set the infalling disc gas, at a given galactocentric radius $R$ 
and time $t$, to have the same abundance pattern as the ISM at $R$ and $t$.
This guarantees the smooth evolution of the gas abundance and of the 
abundance patterns in each radial bin.

\subsection{Disc and halo surface density profiles}

We adopt the following exponential profile for the present-day 
total surface density of the disc component:
\begin{eqnarray}
\Sigma_{t,d}(R,t_{now}) = \Sigma_{t,d}(R_{\odot},t_{now})e^{-(R-R_{\odot})/R_{d}}.
\end{eqnarray}
Here $R_{\odot}=8$~kpc, which is the galactocentric distance of the
Sun within the MW. The same definition of $R_{\odot}=8$ kpc 
is applied to the M31 models.

The surface density profiles of the MW and M31 disc are 
different. We have chosen 
a scale-length of $R_{d} = 3.0$~kpc for the MW
(e.g. Robin, Creze \& Mohan 1992; \citealt{RuphyETal};
\citealt{Freudenreich}) and $R_{d} = 5.5$~kpc for M31
\citep{WK}. For the MW, we assume
$\Sigma_{t,d}(R_{\odot},t_{now}) = 48$~M$_{\odot}$~pc$^{-2}$
\citep{KG}. We adopt $\Sigma_{t,d}(R_{\odot},t_{now}) =
46$~M$_{\odot}$~pc$^{-2}$ in M31, such that the mass of the M31 disc 
is similar to that of the MW disc ($\approx 10^{11}$~M$_{\odot}$ as in
\citealt{Freeman}).

We adopt a modified Hubble law for the present-day total surface
density profile of the halo \citep{BT}:
\begin{eqnarray}
\Sigma_{t,h}(R,t_{now}) = \frac{\Sigma_{t,h_{0}}}{1+(R/R_{\odot})^{\alpha}}.
\end{eqnarray}
Here we set $\alpha = 2$. 
This corresponds to a volume halo density profile:
\begin{eqnarray}
\rho_{t,h}(r,t_{now})=\frac{\rho_{t,h_{0}}}{\Big[1+\big(\frac{r}{R_{\odot}}\big)^{2}\Big]^{3/2}},
\end{eqnarray}
with $\rho_{t,h_{0}} =
2^{3/2}\rho_{t,h}(R_{\odot},t_{now})$ and $\Sigma_{t,h_{0}} =
2R_{\odot}\rho_{t,h_{0}}$. For the MW, we assume
$\Sigma_{t,h}(R_{\odot},t_{now}) = 6$~M$_{\odot}$~pc$^{-2}$ 
which yields a present-day stellar surface density at $R_{\odot}$
of $\Sigma_{\star,h}(R_{\odot},t_{now}) = 1.3$~M$_{\odot}$~pc$^{-2}$
for our model (see Section 3), 
which is consistent with the observed halo stellar density at the solar radius
$\rho_{\star,h}(R_{\odot},t_{now}) = 5.7\times
10^{4}$~M$_{\odot}$~kpc$^{-3}$ as estimated by Preston, Shectman \& Beers
(1991).

The assumption of $\alpha=2$, which implies $\rho_{\star,h}\propto r^{-3}$, 
agrees with a recent analysis
(Zibetti, White \& Brinkmann 2003) of the halo emission for 
a sample of $\approx1000$ edge-on disc galaxies within 
the Sloan Digital Sky Survey (SDSS). 
This result is similar to the density profile 
$\propto r^{-3.5}$ which has been suggested for the MW stellar halo 
(\citealt{CB2000}; \citealt{CB2001}; Sakamoto, Chiba \& Beers 2003).

\subsection{Star Formation Rate}

We assume that the halo star formation (SF) happens on a short
time-scale because of a rapid infall event associated with the
collapse of a single massive proto-galactic cloud \citep{EggenETal} or
multiple mergers of building-blocks \citep{SZ}.  The disc SF is
assumed to be a more quiescent phenomenon, and likely to be driven by the
spiral arms (e.g. \citealt{WS}). Therefore, we adopt a different SF
law for each component.

The adopted halo SFR is described as:
\begin{equation}
\psi_{h}(R,t)=\nu_{h}\Big(\frac{\Sigma_{g}(R,t)}{1~M_{\odot}~pc^{-2}}\Big)^{1.5}
 \,\, {\rm M_{\odot}~Gyr^{-1}~pc^{-2}},
\end{equation}
where $\nu_{h}$ is the star formation efficiency (SFE) in the
halo. Therefore, the halo SFR follows a Schmidt law with exponent
$1.5$ (e.g. \citealt{K}). The adopted halo SFE is $\nu_{h} = 0.125$,
which is approximatively half of the value ($\nu_{h} = 0.25\pm0.07$)
suggested by \cite{K} and is chosen to give the best fit to
the observed halo MDF at the solar neighbourhood (see Section 3). 
Stars born before
$t_{delay}$, when the disc phase starts (Section 2.1), are hereafter
labelled as ``halo stars''.

The adopted disc SFR is written as:
\begin{equation}
\psi_{d}(R,t)=\nu_{d}\Sigma_{g}(R,t)^{2}\,
 \frac{R_{\odot}}{R} \,\, {\rm M_{\odot}~Gyr^{-1}~pc^{-2}},
\end{equation} 
where $\nu_{d}$ is the SFE in the disc. 
This formulation \citep{WS} reflects the
assumption that SF in the disc is triggered by the compression of the
ISM by spiral arms. 
The efficiency
factor $\nu_{d}$ is a free parameter. We have found that $\nu_{d}$
affects both the present-day gas fraction and the disc MDF. 
The value $\nu_{d}=0.03$ is used in our MW model
to reproduce the observed gas density profile of the MW
disc and the observed MDF at the solar neighbourhood. 

\section{The Milky Way model}

Using the multi-zone model described in the previous section, 
we construct a model which most closely reproduces the known
observational properties of the MW. 
The adopted parameters are summarised in Table~1 (MW model).

\begin{figure}
\begin{center}
\includegraphics[width=0.5\textwidth]{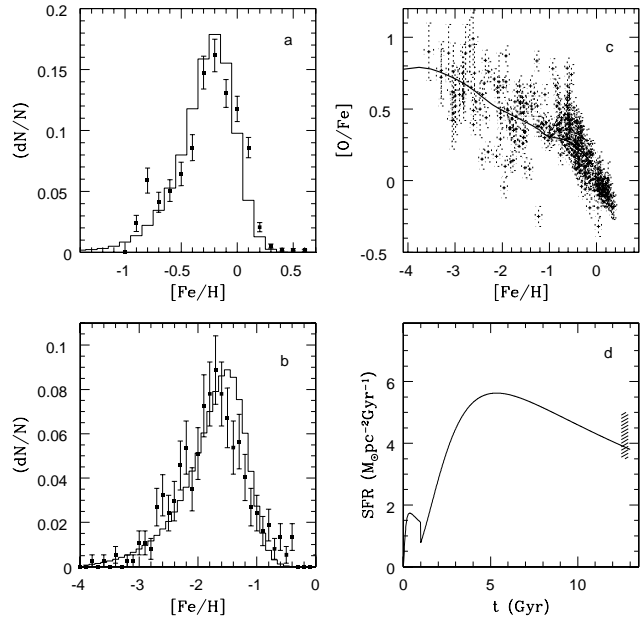}
\caption{ Comparison between the results of MW model (solid lines) 
and observations of a) the MDF in the MW at the solar
neighbourhood (closed boxes with error-bars, Kotoneva~et~al.~2002),
b) halo MDF at $R=R_{\odot}$ (Ryan~\&~Norris~1991, closed boxes with
statistical Poissonian error-bars), c)
[O/Fe] and [Fe/H] for the stars observed at the
solar neighbourhood (Carretta~et~al.~2000, who included reanalysis of
Sneden~et~al.~1991, Tomkin~et~al.~1992, Kraft~et~al.~1992, Edvardsson~et~al.~1993; 
Gratton~et~al.~2003; Bensby,~Feltzing~\&~Lundstr\"om~2003; Cayrel~et~al.~2003), and d)
the present-day SFR at the solar neighbourhood 
as summarised in Rana~(1991, the shaded region).
}
\label{fig1}            
\end{center}
\end{figure}

\begin{figure}
\includegraphics[width=0.5\textwidth]{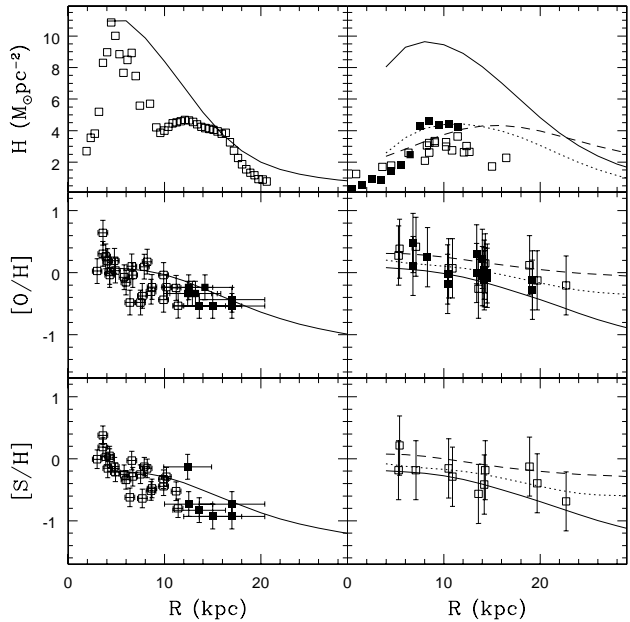}
\caption{Radial distributions of the hydrogen surface density 
(upper) and of the oxygen (middle) and sulfur (bottom) abundances
at the present-day for the MW model (left) and the M31 models (right).
In the left panels, the solid lines are for the MW model results.
In the right panels, solid, dotted and dashed lines display the
M31a, M31b and M31c models, respectively.
The observed distributions of the total surface density of hydrogen
for the MW are from Dame~(1993, open boxes), and those for M31
are from Dame~et~al.~(1993, open boxes) and 
Loinard~et~al.~(1999, closed boxes).
Observations of abundances in HII regions in the MW from Vilchez~\&~Esteban~(1996) 
and Afflerbach~et~al.~(1997) (closed and open boxes,
respectively) are shown in the left panels.
Those in M31 from Dennefeld~\&~Kunth~(1981) and Blair~et~al.~(1982) 
(closed and open boxes, respectively) 
are shown in the right panels.
}
\label{fig2}            
\end{figure}

Fig.~1a compares the model MDF at the solar
neighbourhood with the observed MDF of K dwarfs
(\citealt{KotonevaETal}). Here we have assumed a K dwarf
mass range of 0.8 - 1.4~M$_{\odot}$ and convolved the MDF of the
model with a Gaussian error function with $\sigma=0.15$~dex, 
consistent with the known empirical uncertainties of the observational
data. There is good agreement between the model and the data.
In Fig.~1b, the predicted halo MDF at the solar radius is shown. 
Our MDF includes all stars
still on the main-sequence at the present-time. The MDF has been
convolved with a Gaussian error function with $\sigma=0.25$~dex (e.g. \citealt{A}). 
Our model is in agreement with the observed halo MDF at the
solar neighbourhood \citep{RN}.
The evolution of $[$O/Fe$]$ as a function of $[$Fe/H$]$ for
the gas component in the solar neighbourhood is shown in Fig.~1c. The model result is
consistent with the trend of $[$O/Fe$]$ and $[$Fe/H$]$ observed in local
stars. Fig.~1d displays the predicted SF history at the solar neighbourhood.
The SF history has been estimated by several authors (e.g. \citealt{BN}; \citealt{HernandezETal}).
Unfortunately, such observational estimates are not defined well enough
to provide useful constraints on a chemical evolution model.
Nevertheless, Fig.~1d demonstrates that our MW model is consistent with 
the broad range of the estimated SFR as summarised in \cite{Rana}.

Left panels of Fig.2 show the predicted radial distribution 
of hydrogen column density (upper) and the oxygen (middle)
and sulfur (bottom) abundances of the gas at the present-day, 
and compare them to the observations.
The observed hydrogen surface density is 
obtained by summing the surface densities of H$_{2}$ and HI 
\citep{D}\footnote{The HI surface density profile has been recently 
confirmed by \cite{NS}.}. The result of our MW model is 
compatible with the observed hydrogen distribution at the inner radii 
within the uncertainties ($\approx 50\%$) of the observed values
\footnote{The contribution of HII should also be
considered. However, so far no reliable measurement has been achieved,
since it is not straightforward.}.
However, the hydrogen surface density at the outer radii 
is slightly overestimated when compared with the observations.

The predicted radial abundance profiles of oxygen and sulfur also reproduce
the abundances observed in HII regions 
(\citealt{VE}; \citealt{AfflerbachETal}), although at the outer radii
the model slightly overestimates the sulfur abundances.

Our semi-analytic model, whose parameter values summarised in Table~1,
satisfies the general MW observational constraints.
We now use the same framework to study the chemical evolution of M31.

\section{The M31 models}

The M31 models employ for the disc a different surface density profile  
from that of the MW model. Since the halo density profile of M31 is unknown,
we adopt the same halo profile used in the MW model 
(see Section 2.2).
The SF history of halo and disc is 
described by the parameters listed in Table~1.
In Section~4.1, we first show a M31 model which adopts
the same parameter set of our MW model. This model 
fails to reproduce some crucial features observed in M31.
Therefore, in Sections~4.2 and 4.3, 
we present two models 
able to better explain the key observations.

\subsection{M31a model: MW analogue}

First, we construct a M31 model (M31a) 
with the same parameter values of 
our MW model. These parameters
are summarised in Table~1.
The right panels of Fig.~2 show the radial distributions of 
the hydrogen surface density and
the radial profiles of the oxygen and sulfur abundances of the gas phase. 
The radial profiles of oxygen and sulfur abundances are reproduced within
the observational errors. The M31a model results in too 
high hydrogen surface density when compared with the data.
This is because the observed hydrogen surface density in M31
is smaller than in the MW.

Fig.~3 shows the radial profile for the mean [Fe/H]
($\langle[$Fe/H$]\rangle$) of main-sequence (MS) stars for the M31a model.
Hereafter we simply call $\langle[$Fe/H$]\rangle$ the "mean metallicity".
Model results are compared with
observations by \cite{BellazziniETal}\footnote{The observed MDFs 
are derived from red giant branch (RGB) stars, whereas
the model produces the MDFs of MS stars, 
since it is difficult to construct MDFs of RGB stars within our framework.
We assume that this inconsistency does not affect our comparison
significantly.}.
We have chosen as reference fields those which 
are mostly disc- (or halo-) dominated, with estimated halo- (or disc-) 
contamination around or lower than 10\% 
(see Table~1 in \citealt{BellazziniETal}):
the disc-dominated fields (G287, G119, G33, G76, G322 and G272) 
lie at deprojected\footnote{The inclination angle of the M31
disc is $i_{M31}\approx 12.5^{\circ}$.} galactocentric distances
$R\approx 8$~kpc, $R\approx 12$~kpc, $R\approx 13$~kpc, $R\approx
14$~kpc, $R\approx 15$~kpc and $R\approx 18$~kpc, respectively; the
halo-dominated fields (G319, G11, G351, G219 and G1) lie at 
projected galactocentric distances $R\approx 16$~kpc, $R\approx
17$~kpc, $R\approx 19$~kpc, $R\approx 20$~kpc and $R\approx 34$~kpc,
respectively.
Fig.~3 shows that M31a is in broad agreement 
with the observed mean metallicity in the disc-dominated fields, 
though the metallicity gradient is 
slightly steeper than the observed one (Table~2).
On the other hand, the mean metallicity of the halo in 
M31a is too low, compared to the data. 

To overcome the failure of M31a, we explored the parameter space of the
M31 models which could explain the observational properties of M31,
and found two viable solutions.
In the following, we present these two models.

\begin{figure}
\begin{center}
\includegraphics[width=0.5\textwidth]{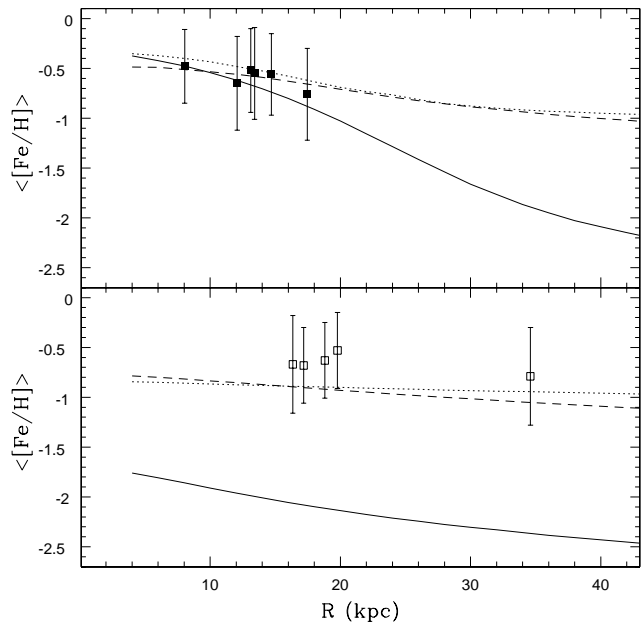}
\caption{ Present-day radial profile of the mean stellar 
metallicity of both disc (upper panel) and halo (lower panel).
Solid, dotted and dashed lines represent the results of the M31a,
M31b and M31c models, respectively.
The mean metallicities from Bellazzini~et~al.~(2003) are also
shown (closed boxes with 1 $\sigma$ dispersion of their MDF).}
\end{center}
\end{figure}

\begin{figure}
\begin{center}
\includegraphics[width=0.5\textwidth]{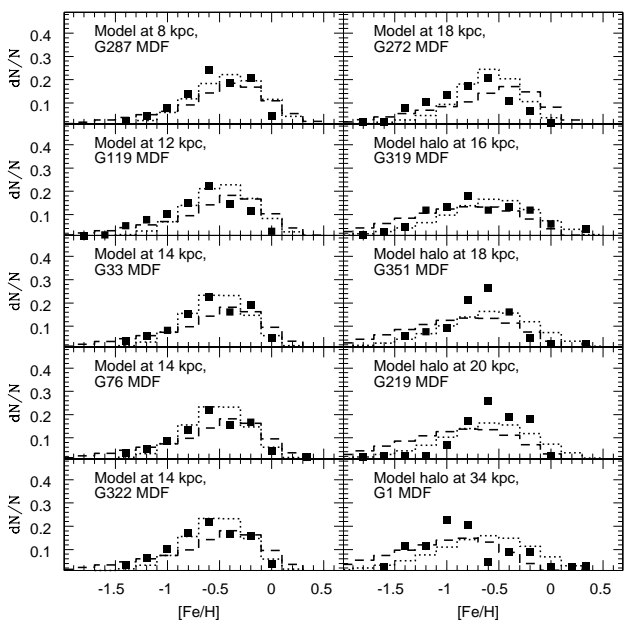}
\caption{The MDFs of the M31b and M31c models (dotted and dashed lines, 
respectively).
The MDFs observed from Bellazzini~et~al.~(2003) are also shown (closed boxes).}
\end{center}
\end{figure}

\subsection{M31b model: our best model}

The model parameters of M31b are summarised in Table~1.
Compared with the M31a model, i.e. the MW analogue, this model has
a shorter time-scale for the infall of disc gas (i.e.
smaller $a_d$ and $b_d$) and higher disc and halo SFE.
We found that a combination of higher disc SFE and 
shorter time-scale for the disc infall
leads to better agreement with the observed 
radial profile of the hydrogen surface density.
In addition, a higher SFE in the halo phase leads to 
a more metal-rich halo, and we adopt a SFE $100$ times higher in the halo
phase than the M31a model.

In M31b the radial profile 
of the oxygen and sulfur abundances reproduces the observational data 
to roughly the same degree of the M31a model
(Fig.~2). On the other hand, the higher disc SFE leads to 
stronger SF and therefore
larger depletion of gas, and lower present-day hydrogen 
surface density than in M31a. 
Consequently, the result of the M31b model
is in better agreement with the observed radial profile of 
the hydrogen surface density (Fig.~2). 
The hydrogen surface density profiles
in both the M31b model and the observations peak at a radius
of $\approx8$~kpc with a value of $\approx$~4.5~M$_{\odot}$~pc$^{-2}$ \citep{LoinardETal}.
However, like in the MW model, the hydrogen surface density at the outer radii 
is overestimated in M31b when compared with observations from \cite{DameETal}. 
We have tried to
reproduce the observed hydrogen surface density at the outer radii by
changing various parameters, however, we are unable to find a better parameter set
than the M31b model. 
This might suggest that, to explain the low gas surface density at the
outer radii in M31, another mechanism which is not included in our semi-analytic
model is required.

The additional benefit of the shorter time-scale of the disc infall 
in M31b is its smaller SFR at the present-day.
Since M31b has a SFR which peaks at an earlier epoch than M31a, 
M31b predicts a lower present-day SFR 
($\approx$~1.5~${\rm M}_{\rm \sun} {\rm yr}^{-1}$) in the disc (4 - 20~kpc)
than M31a ($\approx$~2.4~${\rm M}_{\rm \sun} {\rm yr}^{-1}$).
The observational estimates of the present-day SFR in M31 still 
have a large uncertainty. For example, \cite{W2002}
estimated the SFR for the last $\approx 5$ Gyr between roughly 
$2$ and $20$~M$_{\odot}$~yr$^{-1}$, while \cite{W2003} 
inferred that the total SFR for the
M31 disc has been $\approx 1$~M$_{\odot}$~yr$^{-1}$ over the past $60$~Myr. 
This value is higher than the SFR of $\approx 0.2$~M$_{\odot}$~yr$^{-1}$ 
estimated from H$\alpha$ and Far Infra-Red luminosities 
\citep{DevereuxETal}. 
In addition, all these values could be affected by systematic errors 
(e.g. \citealt{Bell}). Nevertherless, these observations suggest that 
M31 has a lower SFR than the MW, and M31b is consistent with 
this trend.

Fig.~3 shows improved agreement between the results of the
M31b model and the observations of the mean metallicity for both disc- 
and halo-dominated fields. In the disc, M31b has
a shallower metallicity gradient than M31a, and better reproduces
the observed gradient (Table~2). This is mainly due to the small
$b_d$ we adopted (Table~1). 

Fig.~4 directly compares the MDFs of M31b with 
the MDF obtained in \cite{BellazziniETal} in different fields.
The model MDFs are derived from main-sequence stars,
and convolved with a Gaussian error function with $\sigma = 0.25$~dex. 
There is qualitative agreement between the MDFs of the M31b model and 
those observed, especially in the disc-dominated fields. 
The main exception is the G1 halo-dominated field, 
whose observed MDF is more metal-poor and bimodal. 
The MDFs of the G351 and G219 halo-dominated fields
show narrower peaks than those
of the model. Differences in the shape of the observed halo MDFs among
the different fields suggest possible inhomogeneities 
in the M31 halo (see also \citealt{BellazziniETal}).

In the framework of our semi-analytic models,
M31b gives the best fit for the observational constraints
available in M31. In Section~5, 
we discuss the formation process of M31 which is implied by these results. 

\begin{table}
\caption{Radial gradients for the mean stellar metallicity of the MDFs
in the M31 models. 
The results for the MW model and for the fields observed in Bellazzini~et~al.~(2003) 
are also presented as reference.}
\centering
\begin{minipage}{80mm}
\begin{tabular}{@{}lll@{}}
\hline
&$\left(\Delta\langle[Fe/H]\rangle/\Delta R\right)$
\footnote{This value (dex/kpc) refers to the range 4 - 16~kpc for the MW model 
and 4 - 20~kpc for the M31 models.} &
$\left(\Delta\langle[Fe/H]\rangle/\Delta R\right)$
\footnote{This value (dex/kpc) refers to the range 4 - 30~kpc in the halo 
of the MW and M31 models.}\\
\hline
  MW & $-0.058$ & $-0.021~$\\
M31$_{\rm obs}$ & $-0.026$ & $-0.008$\\
M31a & $-0.041$ & $-0.021~$\\
M31b & $-0.021$ & $-0.003~$\\
M31c & $-0.014$ & $-0.009~$\\
\hline
\end{tabular}
\end{minipage}
\end{table}

\subsection{M31c model: another possible model}

The result of the M31b model
suggests that a stronger halo SF
can produce a metal-rich halo.
However, applying the higher SFE is not
the only way to induce stronger SF.
This can also be achieved by employing a
larger reservoir of infalling halo gas, 
without increasing the SFE.
Here we discuss the M31c model,
which assumes a much larger present-day halo surface density
and a longer (6 Gyr) halo phase (Table~1).
Again, to explain the observed hydrogen surface density, M31c uses
a shorter disc infall time-scale, and a weaker radial dependence,
which also leads to a present-day SFR similar to M31b.

Fig.~2. shows the predicted radial distributions of oxygen and sulfur 
abundances which are in agreement with the observational data to roughly 
the same degree of the other models. The hydrogen surface
density profile of the M31c model is consistent with \cite{DameETal}
in the inner region of the disc, $R<12$ kpc. However, M31c 
has a broader peak at a larger radius than M31b.
Consequently, the model overestimates the surface density 
in the outer region, $R>12$ kpc.  
Again, this result might suggest that additional physical mechanisms
operate, 
to explain the observed low hydrogen surface density at the outer radii.

Fig.~3 shows that M31c also reproduces the observed
radial distributions of the metallicity, similarly to M31b.
The metallicity gradient in M31c is flatter than M31b at the inner radii 
as a consequence of the weaker radial dependence
of the disc infall time-scale (i.e. a smaller $b_d$). 
In the halo fields, M31c also has systematically higher metallicity 
than M31a and a steeper gradient than M31b. 
Although the gradient is in better agreement with 
the observations than M31b (Table~2), M31c has lower mean metallicities than observed, 
especially at the outer radii.
In Fig.~4, the predicted MDFs at different radii are 
displayed. The MDFs of M31c are also in good agreement
with those observed, especially in the disc. 
On the other hand, the halo MDFs show a more pronounced
metal-poor tail than in M31b. 

Thus, the M31c model also leads to acceptable results, although 
M31b is in better agreement with the observations. 
The next section discusses the formation history of M31 
implied by the M31b and M31c models.

\section{Discussion}

The results in the previous section have shown 
that the main trends of the chemical evolution
history of both the MW and M31 can be described using the same
framework. 
We have also found that to
explain the observations in more detail, different sets of
parameter values for the formation history are required
for the MW and M31, respectively. 
In this section, we discuss how these different parameters values
relate to different galaxy formation histories.

The previous sections have shown that 
the main differences
between the MW and M31 include 
the hydrogen surface density in the disc and, more significantly, 
the halo metallicity.
The smaller hydrogen surface density in M31
can be explained by a combination of
the shorter disc infall time-scale and the higher disc SFE.

The most striking difference is that 
the mean metallicity observed in the M31 halo is about ten
times higher than that observed in the MW halo, though both
galaxies have similar mass and morphology. \textit{Which is a ``typical''
halo?} Metal-rich halos seem more common than metal-poor ones, as 
pointed out by \cite{ZibettiETal} who analysed a sample
of about 1000 edge-on disc galaxies within the SDSS. 
\cite{HH} have also shown that the NGC5128 halo MDF
closely resembles 
that in the M31 halo.
This similarity
is perhaps suggestive of a common history in the halo formation of both
galaxies, despite their different Hubble-type.

A straightforward way to obtain a more metal-rich halo would 
be by requiring homogeneous pre-enrichment of the infalling halo gas 
with a metallicity $Z\approx 0.1~Z_{\odot}$. 
However, this metallicity seems too high, 
if such pre-enrichment occurred homogeneously in the whole universe.
For example, although this value is close to the mean metallicity
of Damped Lyman $\alpha$ systems (DLAs), 
many DLAs have much lower metallicity
(\citealt{Pettini}; \citealt{ProchaskaETal}). 
If DLAs are more evolved systems than the infalling halo gas, 
their metallicity should be higher than that induced by pre-enrichment. 
Therefore, a pre-enrichment of $Z\approx 0.1~Z_{\odot}$ may not be tenable. 

We found that the observed metal-rich halo of M31 could be explained 
by two scenarios without any pre-enrichment: a) a higher halo SFE; 
b) a larger reservoir of infalling halo gas with a longer halo phase. 
Scenario~a) might be explained in a hierarchical clustering regime, 
where the halo is formed by accretion of building-blocks. 
Theory predicts that gas removal by
supernova-driven winds should be more effective in lower mass
systems, leading to a consequent suppression of the SFR and
a low SFE (e.g. \citealt{DS}; \citealt{E}). 
In addition, observations based on the SDSS also
suggest that the SFE decreases with decreasing stellar mass in low mass
galaxies with a stellar mass M$_{\star}< 3\times10^{10}$~M$_{\odot}$
\citep{KauffmannETal}. Therefore, higher SFE in the M31 halo
phase would be explained if the building-blocks of the M31 halo
are systematically higher mass systems than those of the MW halo
(see also \citealt{HH}).
This result again supports the notion of 
diversity in the formation of spiral galaxies which are apparently
similar in mass and morphology and belong to the same Local Group. 
Admittedly, our
model does not adhere to the hierarchical-clustering
scenario in a self-consistent manner, however, 
our chemical evolution models can be
interpreted as the ``mean'' SF and chemical evolution
history of the stars which end up at each radius.

As an observational consequence of our models,
we found that both scenarios a) and b) predict 
a more massive stellar halo in M31, 
respectively about 6 and 9 times more massive
than in the MW, 
whose stellar halo mass is $\approx 10^{9}$~M$_{\odot}$.
This result suggests 
that there might be a correlation between
the halo metallicity and its stellar mass. 
Using dynamical simulations, Bekki, Harris \& Harris (2003) show that
the stellar halo comes from the outer part of
the progenitor discs when the bulge is formed by a major merger of
two spiral galaxies. Based on this, they also predict the correlation 
of the metallicity of the stellar halos and the mass of the bulges which
were formed by major mergers, since larger bulges have the larger 
progenitors, and progenitor spiral galaxies should follow the observed 
mass-metallicity relation. 
Although they do not mention a correlation between
the masses of the bulge and halo, it would naturally be expected.
Thus, a major merger scenario could explain our conclusion.
Future observational
surveys will better quantify the correlation between the halo
metallicity and its stellar mass (e.g. \citealt{MouhcineETal}).

It is possible to distinguish scenarios~a) and b) through observation.
First, due to a longer halo phase, scenario~b) predicts 
intermediate-age population in the M31 halo. 
This picture could explain the
recent evidence of intermediate-age population
by deep imaging of the M31 halo \citep{BrownETal}.

It is worth mentioning that the metallicity gradient 
in the stellar halo is sensitive to the assumed stellar halo density profile 
especially in scenario~b); a steeper density profile leads to a steeper 
metallicity gradient.
\cite{PvdB1994} suggested that the outer halo of M31 can be modeled 
by a power law surface brightness profile of $I\propto R^{-4}$, 
which is much steeper than what we assumed ($\propto R^{-2}$). 
We found that such a steep profile rules out scenario~b) to
reproduce the flat metallicity gradient observed in the M31 halo.
However, it is still difficult to accurately measure the
halo surface brightness of M31 (e.g. \citealt{ZibettiETal}).
Thus, more observational estimates of the M31 halo 
surface brightness profile would be an important test for this scenario.

We also found that the higher halo SFE in scenario~a) leads to 
about 0.2~dex higher [O/Fe] when [Fe/H]$<$~-1~dex,
due to intense halo SF. Although measuring [O/Fe] is a hard challenge
for the current available instruments, this task could be
accomplished by the next-generation large-aperture optical telescope.

\subsection{Prospect}

The framework we have used can explain the main trends in the chemical
properties of both the MW and M31. However, recent observations
of stellar streams both in the MW 
(e.g. \citealt{HWdZ99}; \citealt{CB2000}; \citealt{ILIC02}; \citealt{BrookETal}; 
\citealt{NHF04}; \citealt{MKL04}) 
and M31 (\citealt{IbataETal};
\citealt{McConnachieETal}; \citealt{MerrettETal};
\citealt{ZuckerETal}; \citealt{LewisETal}) 
clearly identify inhomogeneities in the chemical and
dynamical history of both galaxies, which could suggest that 
a significant
fraction of the halo stars results from later accretion of satellite
galaxies. 
In the light of these recent observations, it will be
important to study both the MW and M31 in more detail by employing 
a self-consistent
chemo-dynamical model (e.g. \citealt{BrookETal2004}) to trace their interrelated chemical and
dynamical histories. 

\section*{Acknowledgments}
We thank the anonymous referee for comments which much improved this manuscript. We acknowledge the financial support of the Australian Research Council through its Discovery Project scheme.

\bsp

\label{lastpage}


\begin{thebibliography}{299}
\bibitem[\protect\citeauthoryear{Afflerbach et al.}{1997}]{AfflerbachETal} Afflerbach A., Churchwell E., Werner M.W., 1997, ApJ, 478, 190
\bibitem[\protect\citeauthoryear{Alib\'es et al.}{2001}]{AlibesETal} Alib\'es A., Labay J., Canal R., 2001, A\&A, 370, 1103
\bibitem[\protect\citeauthoryear{Asplund}{2003}]{A} Asplund A., in ``Elemental Abundances in Old Stars and Damped Lyman $\alpha$ Systems'', 25th meeting of the IAU, Joint Discussion 15, 2003, Sydney, Australia
\bibitem[\protect\citeauthoryear{Bekki \& Chiba}{2001}]{BC} Bekki K., Chiba M., 2001, ApJ, 558, 666
\bibitem[\protect\citeauthoryear{Bekki et al.}{2003}]{BekkiETal} Bekki K., Harris W.E., Harris G.L.H., 2003, MNRAS, 338, 587
\bibitem[\protect\citeauthoryear{Bell}{2003}]{Bell} Bell E.F., 2003, ApJ, 586, 794
\bibitem[\protect\citeauthoryear{Bellazzini et al.}{2003}]{BellazziniETal} Bellazzini M., Cacciari C., Federici L., Fusi Pecci F., Rich M., 2003, A\&A, 405, 867
\bibitem[\protect\citeauthoryear{Bensby et al.}{2004}]{BensbyETal} Bensby T., Feltzing S., Lundstr\"om L., 2004, A\&A, 415, 155
\bibitem[\protect\citeauthoryear{Bertelli \& Nasi}{2001}]{BN} Bertelli G., Nasi E., 2001, AJ, 121, 1013
\bibitem[\protect\citeauthoryear{Binney \& Tremaine}{1987}]{BT} Binney J., Tremaine S., Galactic dynamics, 1987, Princeton University Press, Princeton
\bibitem[\protect\citeauthoryear{Blair et al.}{1982}]{BlairETal} Blair W.P., Kirshner R.P., Chevalier R.A., 1982, ApJ, 254, 50
\bibitem[\protect\citeauthoryear{Brook et al.}{2003}]{BrookETal} Brook C.B., Kawata D., Gibson B.K., Flynn C., 2003, ApJ, 585, L125
\bibitem[\protect\citeauthoryear{Brook et al.}{2004}]{BrookETal2004} Brook C.B., Kawata D., Gibson B.K., Flynn C., 2004, MNRAS, 349, 52
\bibitem[\protect\citeauthoryear{Brown et al.}{2003}]{BrownETal} Brown T.M., Ferguson H.C., Smith E., Kimble R.A., Sweigart A.V. et al., 2003, ApJ, 592, L17
\bibitem[\protect\citeauthoryear{Carretta et al.}{2000}]{CarrettaETal} Carretta E., Gratton R.G., Sneden C., 2000, A\&A, 356, 238
\bibitem[\protect\citeauthoryear{Cayrel et al.}{2003}]{CayrelETal} Cayrel R., Depagne E., Spite M., Hill V., Spite F. et al., 2004, A\&A, 416, 1117
\bibitem[\protect\citeauthoryear{Chang et al.}{1999}]{ChangETal} Chang R.X., Hou J.L., Shu C.G., Fu C.Q., 1999, A\&A, 350, 38
\bibitem[\protect\citeauthoryear{Chiappini et al.}{1997}]{ChiappiniETal} Chiappini C., Matteucci F., Gratton R., 1997, ApJ, 477, 765
\bibitem[\protect\citeauthoryear{Chiba \& Beers}{2000}]{CB2000} Chiba M., Beers T.C., 2000, AJ, 119, 2843
\bibitem[\protect\citeauthoryear{Chiba \& Beers}{2001}]{CB2001} Chiba M., Beers T.C., 2001, ApJ, 549, 325
\bibitem[\protect\citeauthoryear{Cowie \& Songalia}{1998}]{CS} Cowie L., Songaila A., 1998, Nature, 394, 44
\bibitem[\protect\citeauthoryear{Dame}{1993}]{D} Dame T.M., 1993, in Holt S.S., Verter F. eds, ``Back to the Galaxy'', AIP Conf. 278, p. 267
\bibitem[\protect\citeauthoryear{Dame et al.}{1993}]{DameETal} Dame T.M., Koper E., Israel F.P., Thaddeus P., 1993, ApJ, 418, 730
\bibitem[\protect\citeauthoryear{Dekel \& Silk}{1986}]{DS} Dekel A., Silk J., 1986, ApJ, 303, 39
\bibitem[\protect\citeauthoryear{Dennefeld \& Kunth}{1981}]{DK} Dennefeld M., Kunth D., 1981, AJ, 86, 989
\bibitem[\protect\citeauthoryear{Devereux et al.}{1994}]{DevereuxETal} Devereux N.A., Price R., Wells L.A., Duric N., 1994, AJ, 108, 1667
\bibitem[\protect\citeauthoryear{Diaz \& Tosi}{1984}]{DT} Diaz A.I., Tosi M., 1984, MNRAS, 208, 365
\bibitem[\protect\citeauthoryear{Durrell et al.}{2001}]{DurrellETal} Durrell P.R., Harris W.E., Pritchet C.J., 2001, AJ, 121, 2557
\bibitem[\protect\citeauthoryear{Edvardsson et al.}{1993}]{EdvardssonETal} Edvardsson B., Andersen J., Gustafsson B., Lambert D.L., Nissen P.E. et al., 1993, A\&A, 275, 101
\bibitem[\protect\citeauthoryear{Efstathiou}{2000}]{E} Efstathiou G., 2000, MNRAS, 317, 697
\bibitem[\protect\citeauthoryear{Eggen et al.}{1962}]{EggenETal} Eggen O.J., Lynden-Bell D., Sandage A.R., 1962, ApJ, 136, 748
\bibitem[\protect\citeauthoryear{Fenner \& Gibson}{2003}]{FG} Fenner Y., Gibson B.K., 2003, PASA, 20, 189
\bibitem[\protect\citeauthoryear{Ferguson \& Johnson}{2001}]{FJ} Ferguson A.M.N., Johnson R.A., 2001, ApJ, 559, L13
\bibitem[\protect\citeauthoryear{Ferguson et al.}{2002}]{FergusonETal} Ferguson A.M.N., Irwin M.J., Ibata R.A., Lewis G.F., Tanvir N.R., 2002, AJ, 124, 1452
\bibitem[\protect\citeauthoryear{Freeman}{1999}]{Freeman} Freeman K.C., 1999, in Gibson B.K., Axelrod T.S., Putman M.E. eds, ``The Third Stromlo Symposium: The Galactic Halo'', ASP Conference Series, Vol. 165, p. 167
\bibitem[\protect\citeauthoryear{Freeman \& Bland-Hawthorn}{2002}]{FBH} Freeman K., Bland-Hawthorn J., 2002, ARA\&A, 40, 487
\bibitem[\protect\citeauthoryear{Freudenreich}{1998}]{Freudenreich} Freudenreich H.T. 1998, ApJ, 492, 495
\bibitem[\protect\citeauthoryear{Gibson et al.}{2003}]{GibsonETal} Gibson B.K., Fenner Y., Renda A., Kawata D., Lee H.-c., 2003, PASA, 20, 401
\bibitem[\protect\citeauthoryear{Goswami \& Prantzos}{2000}]{GP} Goswami A., Prantzos N., 2000, A\&A, 359, 191
\bibitem[\protect\citeauthoryear{Gratton et al.}{2003}]{GrattonETal2003} Gratton R.G., Carretta E., Claudi R., Lucatello S., Barbieri M., 2003, A\&A, 404, 187
\bibitem[\protect\citeauthoryear{Greggio \& Renzini}{1983}]{GR} Greggio L., Renzini A., 1983, A\&A, 118, 217
\bibitem[\protect\citeauthoryear{Harris \& Harris}{2001}]{HH} Harris W.E., Harris G.L.H., 2001, AJ, 122, 3065
\bibitem[\protect\citeauthoryear{Helmi et al.}{1999}]{HWdZ99} Helmi A., White S.D.M., de Zeeuw P.T., Zhao, H., 1999,  Nature, 402, 53
\bibitem[\protect\citeauthoryear{Hernandez et al.}{2000}]{HernandezETal} Hernandez X., Valls-Gabaud D., Gilmore G., 2000, MNRAS, 316, 605
\bibitem[\protect\citeauthoryear{Holland et al.}{1996}]{HollandETal} Holland S., Fahlman G.G., Richer H.B., 1996, AJ, 112, 1035
\bibitem[\protect\citeauthoryear{Ibata et al.}{2001}]{IbataETal} Ibata R.A., Irwin M.J., Lewis G.F., Ferguson A.M.N., Tanvir N.R., 2001, Nature, 412, 49
\bibitem[\protect\citeauthoryear{Ibata et al.}{2002}]{ILIC02} Ibata R.A., Lewis G.F., Irwin M.J., Cambr\'esy L., 2002, MNRAS, 332, 921
\bibitem[\protect\citeauthoryear{Iwamoto et al.}{1999}]{IwamotoETal} Iwamoto K., Brachwitz F., Nomoto K., Kishimoto N., Umeda H. et al., 1999, ApJS, 125, 439
\bibitem[\protect\citeauthoryear{Kauffmann et al.}{2003}]{KauffmannETal} Kauffmann G., Heckman T.M., White S.D.M., Charlot S., Tremonti C. et al., 2003, MNRAS, 341, 54
\bibitem[\protect\citeauthoryear{Kennicutt}{1998}]{K} Kennicutt R.C. Jr., 1998, ARA\&A, 36, 189
\bibitem[\protect\citeauthoryear{Kotoneva et al.}{2002}]{KotonevaETal} Kotoneva E., Flynn C., Chiappini C., Matteucci F., 2002, MNRAS, 336, 879
\bibitem[\protect\citeauthoryear{Kraft et al.}{1992}]{KraftETal} Kraft R.P., Sneden C., Langer G.E., Prosser C.F., 1992, AJ, 104, 645
\bibitem[\protect\citeauthoryear{Kroupa et al.}{1993}]{KroupaETal} Kroupa P., Tout C.A., Gilmore G., 1993, MNRAS, 262, 545
\bibitem[\protect\citeauthoryear{Kuijken \& Gilmore}{1991}]{KG} Kuijken K., Gilmore G., 1991, ApJ, 367, L9
\bibitem[\protect\citeauthoryear{Lewis et al.}{2004}]{LewisETal} Lewis G.F., Ibata R.A., Chapman S.C., Ferguson A.M.N., McConnachie A.W. et al., in Gibson B.K., Kawata D. eds, Proceedings of Galactic ChemoDynamics V, to appear on PASA 
\bibitem[\protect\citeauthoryear{Loinard et al.}{1999}]{LoinardETal} Loinard L., Dame T.M., Heyer M.H., Lequeux J., Thaddeus P., 1999, A\&A, 351, 1087
\bibitem[\protect\citeauthoryear{Majewski et al.}{2004}]{MKL04} Majewski S.R., Kunkel W.E., Law D.R., Patterson R.J., Polak A.A. et al., 2004, AJ, 128, 245
\bibitem[\protect\citeauthoryear{Matteucci \& Fran\c cois}{1989}]{MF} Matteucci F., Fran\c cois P., 1989, MNRAS, 239, 885
\bibitem[\protect\citeauthoryear{McConnachie et al.}{2003}]{McConnachieETal} McConnachie A.W., Irwin M.J., Ibata R.A., Ferguson A.M.N., Lewis G.F. et al., 2003, MNRAS, 343, 1335
\bibitem[\protect\citeauthoryear{Merrett et al.}{2003}]{MerrettETal} Merrett H.R., Kuijken K., Merrifield M.R., Romanowsky A.J., Douglas N.G. et al., 2003, MNRAS, 346, L62
\bibitem[\protect\citeauthoryear{Moll\'a et al.}{1996}]{MollaETal} Moll\'a M., Ferrini F., Diaz A.I., 1996, ApJ, 466, 668
\bibitem[\protect\citeauthoryear{Mouhcine et al.}{2003}]{MouhcineETal} Mouhcine M., Ferguson H.C., Rich R.M., Brown T., Smith E., 2003, in Combes F., Barret D., and Contini T. eds, ``Semaine de l'Astrophysique Fran\c aise'', Bordeaux, EDP Sciences, Conference Series, p. 122.
\bibitem[\protect\citeauthoryear{Nakanishi \& Sofue}{2003}]{NS} Nakanishi H., Sofue Y., 2003, PASJ, 55, 191
\bibitem[\protect\citeauthoryear{Navarro, Helmi \& Freeman}{2004}]{NHF04} Navarro J.F., Helmi A., Freeman K.C., 2004, ApJ, 601, 43
\bibitem[\protect\citeauthoryear{Pagel \& Tautvai\v siene}{1995}]{PT} Pagel B.E.J., Tautvai\v siene G., 1995, MNRAS, 276, 505
\bibitem[\protect\citeauthoryear{Pettini}{2003}]{Pettini} Pettini M., lectures given at the XIII Canary Islands Winter School of Astrophysics, `Cosmochemistry: The Melting Pot of Elements', Cambridge University Press
\bibitem[\protect\citeauthoryear{Preston et al.}{1991}]{PrestonETal} Preston G.W., Shectman S.A., Beers T.C., 1991, ApJ, 375, 121
\bibitem[\protect\citeauthoryear{Pritchet \& van den Bergh}{1994}]{PvdB1994} Pritchet, C.J., van den Bergh, S., 1994, AJ, 107, 1730
\bibitem[\protect\citeauthoryear{Prochaska et al.}{2003}]{ProchaskaETal} Prochaska J.X., Gawiser E., Wolfe A.M., Castro, S., Djorgovski, S.G., 2003, ApJ, 595, L9
\bibitem[\protect\citeauthoryear{Putman et al.}{2003}]{PutmanETal} Putman M.E., Bland-Hawthorn J., Veilleux S., Gibson B.K., Freeman K.C. et al., 2003, ApJ, 597, 948
\bibitem[\protect\citeauthoryear{Rana}{1991}]{Rana} Rana, N.C., 1991, ARA\&A, 29, 129
\bibitem[\protect\citeauthoryear{Reitzel \& Guhathakurta}{2002}]{RG} Reitzel D.B., Guhathakurta P., 2002, AJ, 124, 234
\bibitem[\protect\citeauthoryear{Reitzel et al.}{2004}]{ReitzelETal} Reitzel D.B., Guhathakurta P., Rich R.M., 2004, AJ, 127, 2133
\bibitem[\protect\citeauthoryear{Renzini \& Voli}{1981}]{RV} Renzini A., Voli M., 1981, A\&A, 94, 175
\bibitem[\protect\citeauthoryear{Rich et al.}{2004}]{RichETal} Rich R.M., Reitzel D.B., Guhathakurta P., Gebhardt K., Ho L.C., 2004, AJ, 127, 2139
\bibitem[\protect\citeauthoryear{Robin et al.}{1992}]{RobinETal} Robin A.C., Creze M., Mohan V., 1992, ApJ, 400, L25
\bibitem[\protect\citeauthoryear{Ruphy et al.}{1996}]{RuphyETal} Ruphy S., Robin A.C., Epchtein N., Copet E., Bertin E. et al., 1996, A\&A, 313, L21
\bibitem[\protect\citeauthoryear{Ryan \& Norris}{1991}]{RN} Ryan S.G., Norris J.E., 1991, AJ, 101, 1865
\bibitem[\protect\citeauthoryear{Sakamoto et al.}{2003}]{SakamotoETal} Sakamoto T., Chiba M., Beers T.C. 2003, A\&A, 397, 899
\bibitem[\protect\citeauthoryear{Sarajedini \& van Duyne}{2001}]{SvD} Sarajedini A., van Duyne J., 2001, AJ, 122, 2444
\bibitem[\protect\citeauthoryear{Schaller et al.}{1992}]{SchallerETal} Schaller G., Schaerer D., Meynet G., Maeder A., 1992, A\&AS, 96, 269
\bibitem[\protect\citeauthoryear{Searle \& Zinn}{1978}]{SZ} Searle L., Zinn R., 1978, ApJ, 225, 357
\bibitem[\protect\citeauthoryear{Sembach et al.}{2002}]{SembachETal} Sembach K.R., Gibson B.K., Fenner Y., Putman M.E., 2002, ApJ, 572, 178
\bibitem[\protect\citeauthoryear{Sneden et al.}{1991}]{SnedenETal} Sneden C., Kraft R.P., Prosser C.F., Langer G.E., 1991, AJ, 102, 2001
\bibitem[\protect\citeauthoryear{Talbot \& Arnett}{1971}]{TA} Talbot R.J. Jr., Arnett W.D., 1971, ApJ, 170, 409
\bibitem[\protect\citeauthoryear{Thilker et al.}{2004}]{ThilkerETal} Thilker D.A., Braun R., Walterbos R.A.M., Corbelli E., Lockman F.J. et al., 2004, ApJ, 601, L39
\bibitem[\protect\citeauthoryear{Timmes et al.}{1995}]{TimmesETal} Timmes F.X., Woosley S.E., Weaver T.A., 1995, ApJS, 98, 617
\bibitem[\protect\citeauthoryear{Tinsley}{1980}]{Tinsley} Tinsley B.M., 1980, Fund. Cosm. Phys., 5, 287
\bibitem[\protect\citeauthoryear{Tomkin et al.}{1992}]{TomkinETal} Tomkin J., Lemke M., Lambert D.L., Sneden C., 1992, AJ, 104, 1568
\bibitem[\protect\citeauthoryear{van den Bergh}{2003}]{vanDenBergh2003} van den Bergh S., in "The Local Group as an Astrophysical Laboratory", Cambridge University Press 2003
\bibitem[\protect\citeauthoryear{Vilchez \& Esteban}{1996}]{VE} Vilchez J.M., Esteban C., 1996, MNRAS, 280, 720
\bibitem[\protect\citeauthoryear{Walterbos \& Kennicutt}{1988}]{WK} Walterbos R.A.M., Kennicutt R.C. Jr., 1988, A\&A, 198, 61
\bibitem[\protect\citeauthoryear{Williams}{2002}]{W2002} Williams B.F., 2002, MNRAS, 331, 293
\bibitem[\protect\citeauthoryear{Williams}{2003}]{W2003} Williams B.F., 2003, AJ, 126, 1312
\bibitem[\protect\citeauthoryear{Woosley \& Weaver}{1995}]{WW} Woosley S.E., Weaver T.A., 1995, ApJS, 101, 181
\bibitem[\protect\citeauthoryear{Worthey \& Espa\~ na}{2003}]{WE} Worthey G., Espa\~ na A.L., 2003, in McWilliam A., Rauch M. eds, Carnegie Observatories Astrophysics Series, Vol. 4: ``Origin and Evolution of the Elements'',  (Pasadena: Carnegie Observatories,\hfill\break\tt http://www.ociw.edu/ociw/symposia/series/\hfill\break symposium4/proceedings.html\rm)
\bibitem[\protect\citeauthoryear{Wyse \& Silk}{1989}]{WS} Wyse R.F.G., Silk J., 1989, ApJ, 339, 700
\bibitem[\protect\citeauthoryear{Zibetti et al.}{2004}]{ZibettiETal} Zibetti S., White S.D.M., Brinkmann J., 2004, MNRAS, 347, 556
\bibitem[\protect\citeauthoryear{Zucker et al.}{2004}]{ZuckerETal} Zucker D.B., Kniazev A.Y., Bell E.F., Martinez-Delgado D., Grebel E.K. et al., 2004, ApJ, 612, L117
\end{thebibliography}
\end{document}